\documentclass[pra,reprint,amsmath,amssymb,superscriptaddress,floatfix]{revtex4-2}

\usepackage{graphicx}
\usepackage{booktabs}
\usepackage{hyperref}
\usepackage{xcolor}
\usepackage{bm}

\newcommand{\Dtwo}{D_{\text{2D}}}
\newcommand{\Dthree}{D_{\text{3D}}}
\newcommand{\Dproj}{D_{\text{proj}}}

\graphicspath{{figures/}}

\begin{document}

\title{Fractal dimension of velocity-driven mixing: Reanalysis of Cao
\textit{et al.}\ (1992) and the universality of Marstrand's codimension rule}

\author{Rod~W.~Douglass}
\email{rwdlanm@gmail.com}
\affiliation{Douglass Research \& Development LLC}

\author{Steven~R.~Dunbar}
\affiliation{Department of Mathematics, University of Nebraska--Lincoln (Professor Emeritus)}

\date{\today}

\begin{abstract}
  Cao \textit{et al.}\ [Phys.\ Rev.\ A \textbf{45}, 3841 (1992)] compared
  box-counting fractal dimensions of interfaces generated by continuum and
  molecular dynamics (MD) simulations of zero-Atwood-number ($A \equiv (\rho_h - \rho_l)/(\rho_h + \rho_l) = 0$) Rayleigh--Taylor-like
  mixing driven by random velocity perturbations rather than by gravitational
  acceleration.
  They reported broadly similar values ($D \sim 1.1$--$1.4$) across physical
  models. We revisit their results with grid-converged geometric
  volume-of-fluid (VOF) simulations and modern box-counting algorithms.
  The grid-converged continuum dimension depends on the
  initial-condition realisation ($D = 1.13$--$1.38$). Re-extraction of
  their MD interface gives $D = 1.373$.  Independent LAMMPS
  simulations (12\,000--500\,000 atoms) show that the MD dimension measures
  thermal scatter of atomic positions, not interface geometry, because
  Cao's box-counting window (box side length $\ell = 0.8$--$2.5\sigma$,
  where $\sigma$ is the Lennard-Jones length scale) lies
  below the scale of the interface structure. A flat interface at
  $t = 0$ gives the same $D \approx 1.36$.  At larger box sizes
  ($\ell = 3$--$50\sigma$), $D$ does rise with time, capturing the
  increasing interface corrugation, but starts at $1.59$ for a flat
  interface --- far above $D = 1.0$ --- because the midpoint cloud's
  thermal thickness contaminates the measurement at all accessible
  scales.  At the atomic scale, the continuum and MD dimensions therefore
  measure different physical quantities --- interface geometry in one
  case, thermal noise in the other.  When the thermal scatter is
  filtered by reducing the midpoint cloud to a one-dimensional curve
  (median binning at $2\sigma$), the MD fractal-dimension trajectory
  converges to the same late-time value ($D \approx 1.43$) as the
  continuum, recovering the cross-model universality that raw
  atomic-scale box counting obscures.

  We prove that the box-counting dimension of the
  linear-regime Rayleigh--Taylor interface is exactly $D = 2 - \beta$ at $A > 0$ and
  $D = \frac{5}{2} - \beta$ at $A = 0$, where $\beta$ is the spectral
  decay exponent of the initial velocity perturbation; the half-unit
  shift is the Lipschitz exponent contributed by the Rayleigh--Taylor
  growth-rate factor $\sqrt{n}$.  For the white-noise initial condition
  used by Cao ($\beta = 0$), both formulas lie outside their domains of
  validity (the underlying random series does not converge), so the
  theorem makes no prediction. The early-time $D \approx 2$ reported by
  Cao \textit{et al.}\ is a box-counting boundary artefact.  The frozen
  nonlinear dimension --- the late-time value reached after the
  initial velocity field has decayed and $D$ ceases to change ---
  depends on the IC realisation (e.g., $D = 1.21$ at $450 \times 450$)
  and records how far nonlinear advection restructures the interface
  from its initial white-noise character.

  We present
  a three-dimensional grid convergence study of the velocity-driven
  mixing interface, obtaining $\Dthree = 2.361 \pm 0.002$ at the finest
  resolution ($400 \times 400 \times 200$). Marstrand's codimension rule
  $\Dthree = \Dtwo + 1$ shows a $7\%$ discrepancy that reflects the
  IC-realisation dependence of $\Dtwo$ at $A = 0$. The prior Rayleigh--Taylor
  results~\cite{Douglass2026a,Douglass2026b} give $2.8$--$6.5\%$ on the same
  unmatched comparison, falling as the three-dimensional grid is refined, so
  what distinguishes $A = 0$ is not the size of the discrepancy but that it
  does not move with resolution. The velocity-stirred case requires matched
  IC realisations for a definitive test.
\end{abstract}

\maketitle

\section{Introduction}
\label{sec:intro}

Cao \textit{et al.}~\cite{Cao1992} compared box-counting fractal dimensions of
interfaces generated by continuum (SOLA-VOF~\cite{HirtNichols1981}) and molecular dynamics (MD)
simulations of zero-Atwood-number ($A \equiv (\rho_h - \rho_l)/(\rho_h + \rho_l) = 0$) Rayleigh--Taylor-like
mixing driven by random velocity perturbations rather than by
gravitational acceleration. They reported $D \sim 1.1$--$1.4$ for the
continuum and $D = 1.357$ for the MD, concluding that the interface
geometry was ``roughly the same'' across physical models --- early evidence
that fractal dimension depends on the mixing \textit{mechanism}, not on
the microscopic physics.

Three decades later, the Cao result warrants revisiting. Their continuum
simulations used grids of $60 \times 60$ to $120 \times 120$ cells --- far
below what is now routine --- and the fractal dimensions they reported
scattered widely with resolution, never reaching a converged value. Their
box-counting algorithm lacked boundary correction, producing an early-time
$D \approx 2$ artefact that they interpreted as a genuine
continuum dimension. And their study was restricted to two
dimensions; no three-dimensional measurement of the velocity-driven mixing
interface was reported.

The missing 3D measurement matters because of a geometric test that
Cao \textit{et al.}\ could not apply without 3D data: Marstrand's
codimension theorem~\cite{Marstrand1954,Mattila1995,Falconer1990}. For a fractal set embedded in
$\mathbb{R}^3$ with isotropic self-similarity, the Hausdorff dimension of a
generic codimension-one slice satisfies $\Dthree = \Dtwo + 1$.
Sreenivasan~\cite{Sreenivasan1991} observed this $+1$ rule empirically for
passive scalar interfaces in turbulent jets and wakes ($\Dtwo \approx 1.36$,
$\Dthree \approx 2.35$), interpreting it as evidence that turbulent scalar
interfaces are isotropically self-similar. More recently, the $+1$ rule has
been shown to hold for buoyancy-driven Rayleigh--Taylor (RT)
interfaces across Atwood numbers from $0.002$ to
$0.60$~\cite{Douglass2026a,Douglass2026b}, to within $0.6$--$4.5\%$ at matched mesh
spacing and $2.8$--$6.5\%$ when each dimensionality is taken at its own
finest grid.  In those cases the instability mechanism
--- bubble competition, merger cascade, secondary Kelvin--Helmholtz
instabilities --- is the same in two and three dimensions.

For velocity-driven mixing, the $+1$ test is non-trivial. In two
dimensions, vortex stretching is geometrically impossible, and the energy
cascade reverses: Kraichnan~\cite{Kraichnan1967} showed that 2D turbulence
exhibits an inverse energy cascade to large scales, in contrast to the
forward energy cascade of 3D turbulence. A 2D simulation of velocity-driven
mixing is therefore not a codimension-one slice of the 3D problem --- it is a
different dynamical system with different spectral transport. Whether the
$+1$ rule survives this dynamical asymmetry has not been tested.

We revisit the Cao problem with geometric VOF simulations on grids up
to $1000 \times 1000$, obtaining a frozen fractal dimension --- the
late-time $D$ reached once the decaying velocity field has ceased to
restructure the interface --- that depends on the initial velocity
realisation. Three controlled
initial-condition (IC) protocols yield converged values of
$D = 1.13$, $1.21$, and $1.38$, demonstrating the absence of a fractal
attractor at $A = 0$ --- in contrast with buoyancy-driven RT, where
$D \approx 1.71$ regardless of IC. Under matched conditions (the
$450^2$ velocity field interpolated onto finer grids), the continuum
dimension converges to $D = 1.21$, well below the re-extracted MD
value $D = 1.37$. Independent LAMMPS~\cite{Thompson2022} simulations
confirm that the MD dimension measures thermal scatter of atomic
positions rather than interface geometry
(Sec.~\ref{sec:universality}), so the two quantities are unrelated.

We prove that the box-counting dimension of the linear-regime RT
interface is exactly $D = 2 - \beta$ at $A > 0$ and
$D = \frac{5}{2} - \beta$ at $A = 0$, where $\beta$ is the spectral
decay exponent of the initial velocity perturbation.  For the
white-noise IC ($\beta = 0$) the series does not converge and the
theorem makes no prediction. The early-time $D \approx 2$ reported by
Cao is a box-counting boundary artefact, not a correct physical
prediction. We present a three-dimensional grid convergence study,
finding $\Dthree \approx 2.36$ at all three resolutions ($150 \times 150 \times 75$
through $400 \times 400 \times 200$). The Marstrand $+1$ test shows a
$7\%$ discrepancy that reflects the IC-realisation dependence of $\Dtwo$
at $A = 0$. On the same unmatched comparison the prior RT results give
$2.8$--$6.5\%$, falling as the three-dimensional grid is refined, and the
passive scalar holds to $1\%$. What distinguishes $A = 0$ is that its
discrepancy is the same at every grid, so the velocity-stirred case requires
matched IC realisations for a definitive test.

\section{Methods}
\label{sec:methods}

\subsection{Continuum simulations}
\label{sec:continuum}

All continuum simulations use OpenFOAM v13 from the OpenFOAM Foundation,
with the incompressibleVoF solver and the multicut MPLIC interface scheme --- a
piecewise-linear interface calculation in the sense of
Ref.~\cite{Youngs1982} --- which computes a piecewise-linear approximation
to the $F = 0.5$ isosurface in each cell. Both fluids have equal density $\rho = 1000$~kg/m$^3$ and kinematic
viscosity $\nu = 10^{-5}$~m$^2$/s, giving $A = 0$ and a Reynolds number
$\text{Re} = V_\text{max} L / \nu = 5000$ based on the peak initial
velocity $V_\text{max} = 0.05$~m/s and domain width $L = 1$~m. Surface
tension is zero. No turbulence model is used.

The two-dimensional simulations are run on a $1.0 \times 1.0$~m square
domain at six resolutions from $150 \times 150$ to $1000 \times 1000$
cells ($\Delta x = 6.67$ to $1.00$~mm). The primary convergence study
uses three grids ($150^2$, $300^2$, $450^2$); three additional grids
($600^2$, $800^2$, $1000^2$) are used in the IC sensitivity study
(Sec.~\ref{sec:ic_sensitivity}). Boundaries
are periodic in $x$ and free-slip on top and bottom. The flat interface is
initialised at $y_0 = 0.5$~m with a sharp VOF field ($F = 1$ below,
$F = 0$ above), where $F(\bm{x},t)$ is the volume fraction of the lower
fluid (the two fluids are identical at $A = 0$; the labelling is
conventional). The velocity is initialised throughout the entire domain
with a band-limited random field generated by FFT; white noise is
transformed to wavenumber space, filtered to retain wavelengths between
$4 \Delta x$ and $L/2$, and transformed back. The short-wavelength cutoff
at $4 \Delta x$ suppresses grid-scale oscillations; the long-wavelength
cutoff prevents domain-scale sloshing. The field is scaled to a peak
magnitude of $V_\text{max}$ and seeded with a fixed random seed for
reproducibility.

The three-dimensional simulations use a $1.0 \times 1.0 \times 0.5$~m
domain at three resolutions: $150 \times 150 \times 75$ ($1.69 \times
10^6$ cells, $\Delta x = 6.67$~mm), $300 \times 300 \times 150$
($13.5 \times 10^6$ cells, $\Delta x = 3.33$~mm), and $400 \times 400
\times 200$ ($32 \times 10^6$ cells, $\Delta x = 2.5$~mm), all with
isotropic spacing. Boundaries are periodic in $x$ and $z$
and free-slip on top and bottom. The velocity initial condition requires
special treatment because a random 3D velocity field is not in general
divergence-free. We generate the initial velocity in three steps:
\begin{enumerate}
  \item Generate three independent realisations of band-limited white noise
        $\hat{u}_i(\bm{k})$ in Fourier space, filtered to $4 \Delta x \leq
        \lambda \leq L_z/2 = 0.25$~m. The upper cutoff is half the
        \emph{shortest} domain dimension, so it is $0.25$~m here against the
        $0.5$~m of the two-dimensional case, whose domain is square.
  \item Apply the Leray projector to enforce $\nabla \cdot \bm{u} = 0$:
        \begin{equation}
          \hat{u}_i^{\text{sol}} = \hat{u}_i
            - k_i \frac{k_j \hat{u}_j}{|\bm{k}|^2},
          \label{eq:leray}
        \end{equation}
        where $\bm{k}$ is the wavevector and summation over $j$ is implied.
        This subtracts the irrotational component of each Fourier mode.
  \item Transform back to physical space and scale to
        $\max|\bm{u}| = V_\text{max}$.
\end{enumerate}
The projection is exact in spectral space: the residual
$\max|\bm{k}\cdot\hat{\bm{u}}|$ is at round-off, as it must be by
construction. That is a check on the projection and not a measurement of the
divergence the solver sees, which is the finite-volume operator applied to
the field after it is interpolated to cell centres and is small but not
zero; the solver removes what remains in its initial \texttt{pcorr}
projection before the first time step. This procedure is the natural 3D
generalisation of a 2D streamfunction, producing a random divergence-free
velocity field with controlled spectral support.

Momentum is advected with a linearUpwind scheme, and the VOF field uses
MPLIC with polynomial sub-cycling. Time integration is first-order Euler
with adaptive time-stepping (maximum Courant number 0.5). Simulations are
run to $t = 50$~s (2D) and $t = 20$--$25$~s (3D, resolution-dependent;
see Sec.~\ref{sec:3d}), with fields written every 0.2~s.

\subsection{Molecular dynamics re-extraction}
\label{sec:md}

Neither the original MD atomic coordinate data nor the MD code used by
Cao \textit{et al.}\ are available. We therefore re-extract the MD
interface geometry from the original publication's figures (this
subsection) and perform independent MD simulations with
LAMMPS (Sec.~\ref{sec:lammps}). Of the two relevant figures, Fig.~2 of
Ref.~\cite{Cao1992} --- which plots the midpoints of the mixing region as
clean dot markers without atomic-scale noise --- provides the higher
fidelity. We render the figure at 600~DPI, detect the axis bounding box,
and extract 703 dot centroids in physical coordinates. Box-counting analysis
of these midpoints (Sec.~\ref{sec:boxcounting}) yields $D = 1.373$ with
$R^2 = 0.984$ over 1.41 decades of scaling, within 0.016 of the published
value of 1.357. The small difference is attributable to the boundary correction
and fitting algorithm used here.

We also extract the interface from Fig.~1(a), which shows the full MD
snapshot including atoms. After skeletonisation (morphological thinning
of the rasterised boundary to a one-pixel-wide curve), box counting gives
$D = 1.30$, lower than the Fig.~2 value by 0.07 due to raster
line-width artefacts. The Fig.~2 extraction, being based on centroid
positions rather than pixel boundaries, is the more reliable estimate and
is used throughout.

\subsection{LAMMPS molecular dynamics simulations}
\label{sec:lammps}

To obtain a fresh set of two- and three-dimensional MD data for
comparison with the continuum results, and to determine what
the MD box-counting dimension actually measures, we performed
independent simulations using LAMMPS~\cite{Thompson2022}.  The setup reproduces the essential
features of Cao \textit{et al.}'s configuration: a two-dimensional
rectangular domain with 2:1 aspect ratio, Lennard-Jones 6-12
interactions ($\varepsilon = \sigma = 1$ in reduced units, cutoff
$r_c = 2.5\sigma$), atomic mass $m = 16$~amu (Cao's choice, lighter
than argon to accelerate the dynamics), and heavy frozen wall atoms
around the perimeter.  Fluid atoms are placed on a triangular lattice
at number density $\rho = 0.85\,\sigma^{-2}$, with the bottom half
labelled type~1 and the top half type~2.  At $A = 0$ the two species
are identical (same mass, same potential), so the single-box
initialisation is physically equivalent to Cao's two-half-cell
procedure.

\begin{figure}[b]
  \includegraphics[width=\columnwidth]{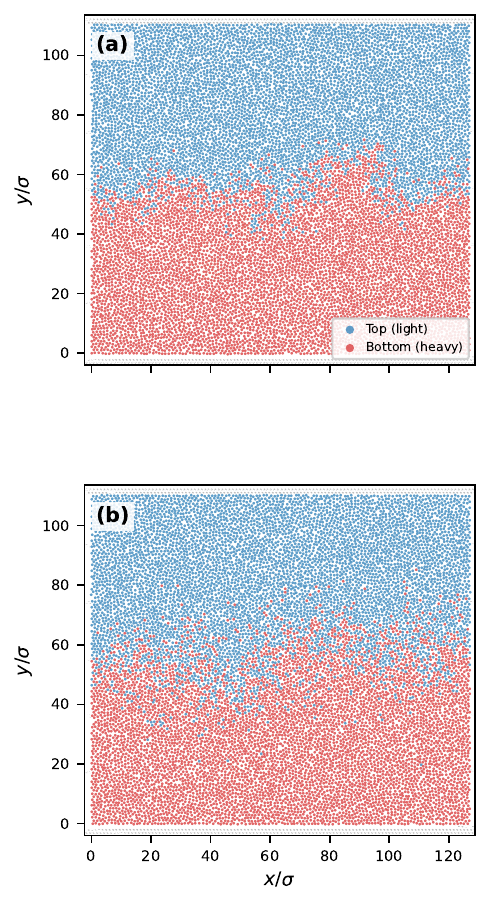}
  \caption{\label{fig:md_snapshot}%
    LAMMPS atom positions for the $12\,000$-atom system at
    $T^* = 2.5$.  Red: bottom-half atoms (type~1, ``heavy''); blue:
    top-half atoms (type~2, ``light''); grey: frozen wall atoms.
    (At $A = 0$ the two species have equal mass; the labels are
    conventional.)
    (a)~$t^* = 0$ (reduced time $t^* = t/\tau_\mathrm{LJ}$): flat
    interface with thermal jitter.
    (b)~$t^* = 198$: velocity-driven interpenetration has broadened
    the mixing zone, but the midpoint cloud retains the same
    atomic-scale box-counting dimension $D \approx 1.36$
    (Sec.~\ref{sec:universality}).}
\end{figure}

Initial velocities are drawn from a Maxwell--Boltzmann distribution at
reduced temperature $T^* = k_B T / \varepsilon = 2.5$ ($\approx 300$~K
for argon-like LJ), where $k_B$ is Boltzmann's constant, followed by
NVT (constant particle number, volume, and temperature;
Nos\'{e}--Hoover thermostat) equilibration at timestep
$\Delta t = 0.002\,\tau_\mathrm{LJ}$, where the LJ time unit is
$\tau_\mathrm{LJ} = \sigma\sqrt{m/\varepsilon}$ --- $50\,000$ steps for
the smallest system and $100\,000$ for the four larger ones, the longer
equilibration being needed to melt the bigger lattices.  Reduced time is
$t^* = t / \tau_\mathrm{LJ}$.  Note that the simulations are run in reduced
Lennard-Jones units, in which the fluid atom mass is unity; Cao
\textit{et al.}'s $16$~a.u.\ enters only in converting $t^*$ to physical
time and plays no part in the dynamics reported here.  Five system sizes are
used, from $12\,000$ to $500\,000$ fluid atoms ($11\,881$, $25\,289$,
$50\,065$, $100\,618$ and $503\,821$), and temperatures of $T^* = 1.0$,
$2.5$ and $5.0$ are tested at the intermediate size.  On the Ar-Ar mapping
($\varepsilon/k_B \approx 120$~K) these correspond to $120$, $300$ and
$600$~K, bracketing Cao's $130$--$900$~K range from below at the hot end.  The interface is defined by the
midpoint convention of Cao --- for each column of atoms, the midpoint
between the lowest type-1 and highest type-2 atom.
Figure~\ref{fig:md_snapshot} shows the initial and late-time atom
distributions for the $12\,000$-atom case.

\subsection{Box-counting algorithm}
\label{sec:boxcounting}

Throughout this paper, $D$ denotes the box-counting dimension
(Minkowski--Bouligand dimension), computed with three refinements
over the algorithm used by Cao \textit{et al.}; full details and validation
are given in Ref.~\cite{Douglass2025}.

\textit{Boundary removal.} Let $\ell$ denote the box side length and
$N(\ell)$ the number of boxes of side $\ell$ that intersect the interface.
At the largest $\ell$, $N(\ell)$ is
depressed because the interface does not fill the domain; at the smallest
$\ell$, $N(\ell)$ may saturate at the grid or pixel scale. We detect
these boundary effects by comparing the local slope of the log-log curve
in the first, middle, and last segments, and trim endpoints whose slope
deviates significantly from the interior. This eliminates the early-time
$D \approx 2$ artefact in Cao's data, which arose from uncorrected
large-$\ell$ contamination when the interface was nearly flat.

\textit{Optimal scaling region.} After boundary removal, we identify the
best scaling window by sliding a window of at least 5 points along the
log-log curve and selecting the interval that maximises $R^2 \times
\Delta \log \ell$, subject to $R^2 \geq 0.97$ and
$\Delta \log \ell \geq 1.0$ decade.

\textit{Multi-offset averaging.} To reduce lattice bias, box counts are
computed over 8 (2D) or 8--27 (3D) translational offsets of the
box grid, and the minimum count at each box size is retained. The
offsets use covering placement ($x_\text{origin} = x_\text{min} - f\ell$,
$f = 0, 0.25, 0.50, 0.75$), ensuring that every grid covers the full
interface extent. The maximum box size is $L/8$, where $L$ is the domain
side length.

\textit{Scaling-range limitations.} The 3D box-counting has only
${\sim}1$ decade of usable scaling range, limited by the narrow mixing
layer (${\sim}5$--$9$~cm) in the $1$~m domain, compared with
${\sim}1.5$ decades in 2D. With only 5--6 points in the log-log
regression, $R^2$ values peak near $0.998$; no timestep at any
resolution reaches $R^2 \geq 0.999$. This limited scaling range
produces visible scatter in the per-snapshot $\Dthree$ values
[Fig.~\ref{fig:D3d}(b)], particularly at early times when
the mixing layer is thin. The same effect is present in the 2D
measurements at early times [Fig.~\ref{fig:D3d}(a)], where
the interface is nearly flat and the scaling range is similarly
compressed. The frozen-window averages used for the codimension
test (Table~\ref{tab:3d_convergence}) are taken from late times when the
interface is well-developed and the snapshot-to-snapshot scatter is
small ($\sigma_D \leq 0.03$).

For the 2D continuum simulations, the interface is the $F = 0.5$
contour extracted from the VOF field. For the 3D simulations, the
$F = 0.5$ isosurface is extracted as a triangulated mesh using
OpenFOAM's \texttt{isoSurface} function object with cell-point
interpolation; box counting is performed on the triangle set. Projected
dimensions $\Dproj$ are computed by projecting the 3D isosurface onto a
coordinate plane and performing 2D box counting on the resulting
silhouette. The values quoted here collapse the $z$ axis, giving the
$xy$-plane shadow. The choice matters, because the interface is a sheet
whose normal is $y$: collapsing $z$ views it edge-on and gives
$\Dproj = 1.456$, while collapsing $y$ views it face-on and gives $1.82$
at the same largest box. Neither is compared with a measured counterpart
below.

\section{Theoretical bounds on the interface dimension}
\label{sec:bounds}

Before presenting the numerical results, we derive exact bounds on
the fractal dimension of the interface in the linear regime.  The
derivation is two-dimensional: the interface is a graph
$\zeta'(x_1,t)$ over a single horizontal coordinate (here and
throughout this section, primes denote perturbations from the flat
base state, not derivatives), so the resulting dimension $D$ lies in
$[1,2]$.  Subject to that restriction,
the derivation applies to any Rayleigh--Taylor configuration with zero
surface tension; we specialise to the $A = 0$ velocity-driven case at
the end of this section.

\subsection{Interface displacement and the velocity spectrum}
\label{sec:displacement}

In the zero-surface-tension case ($\mathcal{T} = 0$), a normal mode
analysis of the linearised equations~\cite{BellmanPennington1954} gives
a dispersion relation $s_n^2 = A g k_n$, where
$k_n = 2\pi n/L$ is the wavenumber of mode $n$.  The temporal growth
rate is therefore $s_n \propto \sqrt{n}$, and the interface displacement
can be written
\begin{equation}
    \zeta'(x_1,t) = \sum_{n=1}^{\infty} a_n e^{\sqrt{n}\,t}
    \cos(2\pi n x_1/L),
\label{eq:zeta}
\end{equation}
where $a_n$ is the initial amplitude of mode $n$ and finite-domain
corrections are negligible ($\tanh(n\pi) \approx 1$ for all $n \geq
1$).  Differentiating with respect to time and evaluating at $t = 0$
gives the initial interfacial velocity
\begin{equation}
    w'(x_1,0,0) = \sum_{n=1}^{\infty} b_n \cos(2\pi n x_1/L),
\label{eq:w}
\end{equation}
where $b_n \equiv a_n\sqrt{n}$.  The fractal dimension of the
interface graph is determined by the decay rate of these velocity
coefficients.

\subsection{Upper bound}
\label{sec:upper}

Assume that the random velocity coefficients have the form $b_n =
X_n/n^\beta$, where $0 \le \beta < 1$ and the $X_n$ are independent
and uniformly distributed on $[-1,1]$.  Independence and symmetry of
the $X_n$ place the series in the class treated by
Kahane~\cite{Kahane1985}, and his regularity condition
$\mathbb{E}(X_n^4) \le C\,\mathbb{E}^2(X_n^2)$, where $\mathbb{E}$ denotes the ensemble expectation, (eq.~(4), p.~93), which
both theorems below require, holds with $C = 9/5$ for every $n$.  The
exponent $\beta$ controls the spectral decay:
$\beta = 0$ corresponds to white noise~\cite{Buckingham1983},
$\beta = 1/2$ to $1/f$~noise, and $\beta \to 1$ to Brownian motion.
Then $a_n = b_n/\sqrt{n} \propto 1/n^{\beta + 1/2}$.  Define the
dyadic partial-variance sums
\begin{equation}
    c_j = \biggl(\sum_{2^j \le n < 2^{j+1}}
    \mathbb{E}(a_n^2)\biggr)^{1/2}
    = O(2^{-\beta j}),
\end{equation}
where the asymptotic form follows from the Euler summation formula.
By Theorem~3 (p.~89) of Kahane~\cite{Kahane1985}, the interface
profile $\zeta'(x_1)$ belongs to the Lipschitz class
$\Lambda_\alpha$ for every $\alpha < \beta$.  The
Besicovitch--Ursell bound (Corollary~11.2(a), p.~147 of
Falconer~\cite{Falconer1990}) then gives
\begin{equation}
    D \le 2 - \alpha \quad \text{for all}\;\alpha < \beta,
    \quad\text{hence}\quad D \le 2 - \beta.
\label{eq:upper}
\end{equation}

\subsection{Lower bound}
\label{sec:lower}

A matching lower bound follows from the irregularity of $\zeta'$.
With $b_n \propto 1/n^\beta$ so that $a_n \propto 1/n^{\beta+1/2}$,
the condition $\liminf_{n\to\infty} n^{1+2\gamma}\,\mathbb{E}(a_n^2)
> 0$ is satisfied for every $\gamma \ge \beta$.  Theorem~4 (p.~101)
of Kahane~\cite{Kahane1985}, which is stated for $0 < \gamma < 1$, then
guarantees, almost surely, that
\begin{equation}
    \limsup_{h \to 0}
    \frac{|\zeta'(x_1+h) - \zeta'(x_1)|}{|h|^\gamma} > 0
\label{eq:irreg}
\end{equation}
for every $x_1$ and every $\gamma$ with $\beta \le \gamma < 1$.  Since
$\beta < 1$ that range is non-empty, which is all the argument needs.
Covering the graph with
boxes of side $\ell$ and using~(\ref{eq:irreg}), each vertical
column of width $\ell$ requires at least
$\ell^{\gamma-1}$ boxes, giving a total box count
$N(\ell) \ge \ell^{\gamma - 2}$.  Since this holds for
any $\gamma > \beta$,
\begin{equation}
    D \ge 2 - \gamma \quad\text{for all}\;\gamma > \beta,
    \quad\text{hence}\quad D \ge 2 - \beta.
\label{eq:lower}
\end{equation}

\subsection{The result}
\label{sec:Dexact}

Combining~(\ref{eq:upper}) and~(\ref{eq:lower}):
\begin{equation}
    \boxed{D = 2 - \beta}
\label{eq:Dbeta}
\end{equation}
almost surely.  The fractal dimension of the linear-regime interface
is determined exactly by the spectral decay exponent of the initial
velocity perturbation.  The same exponent follows independently from
the spectral estimate for the dimension of a graph.  Since
$\mathbb{E}(a_n^2) \propto n^{-(1+2\beta)}$, the profile has a
$1/\omega^{p}$ power spectrum with $p = 1 + 2\beta$, and the
estimate $\dim_\mathrm{B} = \tfrac{1}{2}(5 - p)$
(eqs.~(11.20)--(11.21), p.~158 of Falconer~\cite{Falconer1990}, where this
exponent is written $\alpha$)
returns $2 - \beta$.  That route is a heuristic where the two bounds
above are not, so it corroborates the result rather than establishing
it.

\subsection{Application to the velocity-driven problem}
\label{sec:apply}

Equation~(\ref{eq:Dbeta}) was derived for the \textit{displacement}
graph $\zeta'(x_1)$, whose coefficients $a_n = b_n/\sqrt{n}$ decay
faster than the velocity coefficients $b_n$ by the factor $\sqrt{n}$
arising from the RT growth rate.  At $A = 0$ the growth factors
in~(\ref{eq:zeta}) are absent: the interface displacement $\zeta'(x_1,t)
\approx w'(x_1,0,0)\cdot t$ inherits the velocity coefficients $b_n$
directly, without the $\sqrt{n}$ regularisation.  Repeating the
derivation of Secs.~\ref{sec:upper}--\ref{sec:lower} with $b_n$ in
place of~$a_n$, the dyadic partial-variance sums become
\begin{equation}
    c_j = \biggl(\sum_{2^j \le n < 2^{j+1}}
    \mathbb{E}(b_n^2)\biggr)^{1/2}
    = O(2^{\,(1/2 - \beta)\,j}),
\end{equation}
which decay only for $\beta > 1/2$.  Kahane's upper and lower bounds
then give
\begin{equation}
    D = \tfrac{5}{2} - \beta
    \qquad (\tfrac{1}{2} < \beta < \tfrac{3}{2}).
\label{eq:Dbeta_A0}
\end{equation}
The shift from $2 - \beta$ to $\frac{5}{2} - \beta$ is exactly the
half-unit of Lipschitz exponent contributed by the RT growth-rate
factor~$\sqrt{n}$.  The spectral estimate confirms~(\ref{eq:Dbeta_A0}):
$\mathbb{E}(b_n^2) \propto n^{-2\beta}$ gives a power-spectrum exponent
$p = 2\beta$ and $\dim_\mathrm{B} = \tfrac{1}{2}(5 - p) =
\tfrac{5}{2} - \beta$.

For $\beta \le 1/2$ --- including the white-noise case $\beta = 0$ used
in the production simulations --- the Lipschitz bounds are vacuous and
$\sum \mathbb{E}(b_n^2)$ diverges, so the random series has no $L^2$
limit and the theorem makes no prediction.  (The same convergence
failure affects the $A > 0$ formula at $\beta = 0$, where $\sum
\mathbb{E}(a_n^2) \sim \sum 1/n$ diverges logarithmically.)  The
defensible statement at $\beta = 0$ concerns the band-limited partial
sums that the simulations actually contain: a finite sum of $N$ cosines
with flat amplitudes is well-defined, and its graph becomes increasingly
space-filling as $N$ grows.  The measured $D$ therefore rises toward~$2$
with grid refinement, a property of the sequence of partial sums rather
than of a limit function.

In practice, the band limitation restricts the IC to $N_\text{modes}
\sim N_x/8$ to $N_x/2$ Fourier modes.  A finite sum of cosines is
smooth below the shortest wavelength, so the measured $D$ is pulled
below the asymptotic prediction.  The frozen dimension $D = 1.21$
(Sec.~\ref{sec:convergence}) reflects further reduction by nonlinear
advection, which restructures the interface spectrum over the full
simulation duration.

\subsection{Numerical test with amplitude-per-mode initial conditions}
\label{sec:betatest}

To test the $\beta$-dependence numerically, we run nine additional cases
at $\beta = 0.25$, $0.50$, and $0.75$ on three grids ($300^2$, $600^2$,
$1200^2$; these grids differ from the production set in
Sec.~\ref{sec:continuum} because the $\beta$-test requires
finer resolution to separate the spectral slopes), using an
amplitude-per-mode normalisation that preserves
the energy in each Fourier mode as the grid is refined (phases are
random, amplitudes decay deterministically as $|k|^{-\beta}$).  A
deterministic modulus carrying a uniform random phase is still an
independent symmetric coefficient in the sense of
Sec.~\ref{sec:upper}, and it meets the regularity condition there with
$C = 1$, so the derivation applies to these initial conditions
unchanged.  Because the $\beta$-test cases are at $A = 0$, the
relevant prediction is~(\ref{eq:Dbeta_A0}), $D = \frac{5}{2} - \beta$,
not~(\ref{eq:Dbeta}).  Of the three test values, only $\beta = 0.75$
lies in the theorem's domain ($\beta > 1/2$); at $\beta = 0.25$ and
$0.50$ the series does not converge and the theorem makes no
prediction.  Both formulas share the same $\beta$-sensitivity:
$\partial D/\partial\beta = -1$.

\begin{figure}
  \includegraphics[width=\columnwidth]{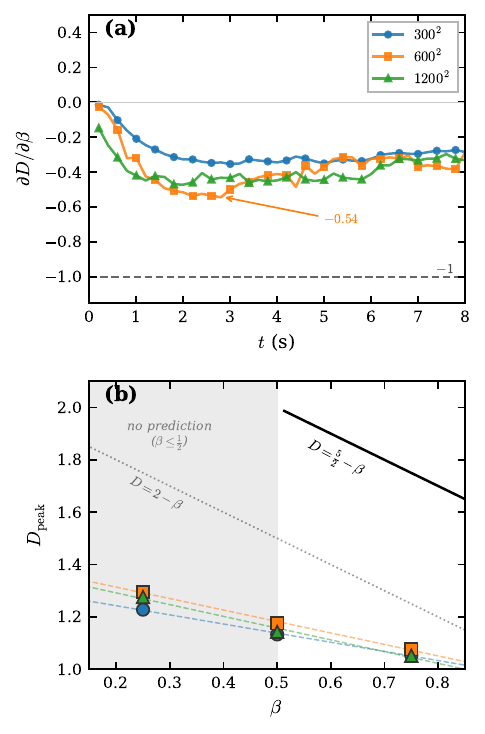}
  \caption{Numerical test of the $\beta$-dependence at $A = 0$.
    (a)~Instantaneous slope $\partial D/\partial\beta$ vs time for three
    grids with amplitude-per-mode initial conditions.  Dashed line:
    predicted $-1$ (common to both $2 - \beta$ and
    $\frac{5}{2} - \beta$).  The $600^2$ grid reaches $-0.54$ at
    $t \approx 2.8$~s; all grids settle to a nonlinear plateau near
    $-0.35$ by $t \approx 4$~s.
    (b)~Peak early-time dimension $D_\text{peak}$ vs $\beta$; coloured
    symbols as in~(a).  Shaded region: $\beta \le 1/2$, where the
    series does not converge and the theorem makes no prediction.
    The measured slopes ($\approx 44\%$ of the predicted~$-1$) and
    absolute levels ($D_\text{peak} \approx 1.05$ at $\beta = 0.75$)
    reflect the rapid onset of nonlinear dynamics.}
  \label{fig:beta_slope}
\end{figure}

Figure~\ref{fig:beta_slope}(b) shows the early-time peak dimension
$D_\text{peak}$ (maximum of $D(t)$ for $t \le 5$~s) at each $\beta$
and grid, together with the theoretical lines $D = \frac{5}{2} - \beta$
(for $\beta > 1/2$) and $D = 2 - \beta$.  The measured $D_\text{peak}$
decreases with $\beta$ at every resolution, with a fitted slope
$\partial D_\text{peak}/\partial\beta$ of $-0.35$ ($300^2$), $-0.44$
($600^2$) and $-0.45$ ($1200^2$) --- about $44\%$ of the
predicted~$-1$.

Figure~\ref{fig:beta_slope}(a) tracks the instantaneous slope
$\partial D / \partial\beta$ as a function of time.  The $600^2$ grid
reaches a peak slope of $-0.54$ at $t \approx 2.8$~s (54\% of the
linear prediction), before nonlinear mode interactions pull all three
grids back to a plateau near $-0.35$ by $t \approx 4$~s.  The finer
$1200^2$ grid transitions to nonlinearity faster --- its peak comes at
$t \approx 2.0$~s rather than $2.8$~s --- and reaches only $-0.47$.

The absolute $D_\text{peak}$ at $\beta = 0.75$ --- the only test
point inside the theorem's domain --- is $1.05$--$1.08$, well below
both predictions ($1.25$ for $2 - \beta$, $1.75$ for
$\frac{5}{2} - \beta$).  More fundamentally, the $\beta$-test
interfaces present as smooth curves: at $\beta = 0.75$ on the
$1200^2$ grid, local log-log slopes rise to $\sim\!1.25$ in the
middle of the box-size range but return to $\sim\!1.0$ at the
small-box end, where box dimension is defined, and refining from
$300^2$ to $1200^2$ extends the range with points that all arrive
at slope~$\approx 1$ (mean of three smallest-box slopes: $1.118$,
$1.051$, $1.008$).  A Koch curve ($D = 1.262$) measured through the
same pipeline oscillates about its true dimension with no drift
toward~$1$.  A smooth sine ($D = 1$ exactly) box-counted over the
same window returns fitted ``dimensions'' of $0.93$--$1.84$
depending on amplitude and wavelength (sub-unity values are a
finite-segment edge effect); the upper end exceeds the
$\frac{5}{2} - \beta$ prediction of $1.75$ at $\beta = 0.75$ using a
curve with no fractality in it, and the $\beta$-test values sit
squarely in this range.  The $\beta$-test demonstrates that $D$
responds to the spectral exponent (the ordering is robust across
grids), but the interfaces do not develop fractal structure resolvable
by box counting, so neither intercept can be tested numerically.

\section{Results}
\label{sec:results}

\subsection{Grid convergence and the early-time artefact}
\label{sec:convergence}

Table~\ref{tab:convergence} summarises the 2D grid convergence of the
frozen fractal dimension. Both MPLIC and the original SOLA-VOF code of Cao
\textit{et al.}\ are included; the SOLA-VOF runs use the same grid series
with a 5-point box-average filter for the initial velocity instead of the
FFT band-pass, reproducing the original code's methodology at modern
resolution.

\begin{table}[b]
  \caption{\label{tab:convergence}%
    Frozen fractal dimension $D$ for the 2D velocity-driven mixing
    interface ($A = 0$, $V_\text{max} = 0.05$~m/s), averaged over
    $t = 40$--$50$~s.  Each resolution uses an independent IC realisation
    (Sec.~\ref{sec:ic_sensitivity}), so the MPLIC column conflates grid
    refinement with IC change.}
  \begin{ruledtabular}
  \begin{tabular}{lccc}
    Resolution & $\Delta x$ (mm) & $D$ (MPLIC) & $D$ (SVOF) \\
    \hline
    $150 \times 150$  & 6.67 & $1.31 \pm 0.01$ & $1.32 \pm 0.01$ \\
    $300 \times 300$  & 3.33 & $1.24 \pm 0.03$ & $1.31 \pm 0.01$ \\
    $450 \times 450$  & 2.22 & $1.205 \pm 0.006$ & $1.29 \pm 0.01$ \\
    $600 \times 600$  & 1.67 & $1.160 \pm 0.010$ & --- \\
    $800 \times 800$  & 1.25 & $1.17 \pm 0.01$ & --- \\
    $1000 \times 1000$ & 1.00 & $1.131 \pm 0.001$ & --- \\
  \end{tabular}
  \end{ruledtabular}
\end{table}

The MPLIC results decrease from $1.31$ at $150 \times 150$ to $1.21$
at $450 \times 450$, then drop further to $1.16$ at $600^2$ before
settling at $1.17$--$1.13$ on the two finest grids.
The SVOF values are less regular, reflecting the greater numerical
diffusion of the donor-acceptor reconstruction; at $450 \times 450$
the SVOF interface has not fully frozen by $t = 50$~s, and the
late-time $D = 1.29$ still carries a slow upward drift.  As shown in
Sec.~\ref{sec:ic_sensitivity}, this decrease conflates grid refinement
with a change in the IC realisation; the $450^2$ value $D = 1.21$ is
the frozen dimension for one particular velocity field, not a
universal grid-converged limit.

Cao \textit{et al.}\ reported $D \approx 2$ at early times ($t < 5$~s),
interpreting this as a physical difference between continuum and atomistic
interfaces.  As shown in Sec.~\ref{sec:apply}, the white-noise IC
($\beta = 0$) lies outside the theorem's domain of validity at $A = 0$
(the random series does not converge), so the theorem makes no prediction
at $\beta = 0$.  The Cao measurement was dominated by a box-counting
boundary artefact: the early-time interface is nearly flat --- a 1D object
embedded in a 2D domain --- and the box counts at the largest $\ell$
reflect the rectangular domain boundary rather than interface structure.
Our boundary-removal algorithm (Sec.~\ref{sec:boxcounting}) strips this
contamination.  Once the deviant large-scale points are trimmed, the
early-time $D$ begins near 1.0 (a flat line) and rises as the velocity
field wrinkles the interface, but reaches only $D \approx 1.0$--$1.2$ at
the resolved grids ($300^2$ and $450^2$).  The $D \approx 2$ that Cao
reported was thus a box-counting boundary artefact; removing it eliminates
the distinction Cao drew between continuum and atomistic early-time
behaviour.

The re-extracted MD dimension is $D = 1.373$
(Sec.~\ref{sec:md}), above the continuum value.
Independent LAMMPS simulations at system sizes from $12\,000$ to
$500\,000$ atoms confirm this value but reveal that it measures thermal
scatter of atomic positions rather than interface corrugation; the full
MD analysis is deferred to Sec.~\ref{sec:universality}.  Both interfaces
are monofractal; the generalised dimensions satisfy $D_0 \approx D_1
\approx D_2$ to within measurement uncertainty for the continuum data.

\begin{figure}[t]
  \includegraphics[width=\columnwidth]{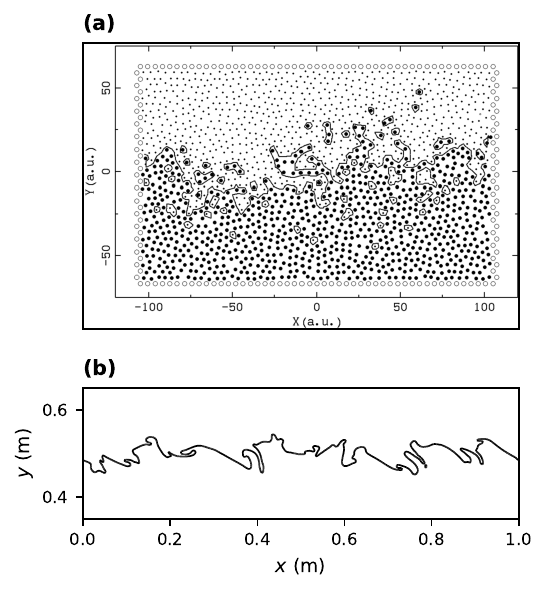}
  \caption{\label{fig:md_continuum}%
    Interfaces from (a)~the molecular dynamics simulation of Cao
    \textit{et al.}~\cite{Cao1992} (their Fig.~2a; reproduced with
    permission from the American Physical Society) and (b)~the present
    MPLIC $450 \times 450$ continuum simulation at $t = 50$~s (frozen
    state). The initial interface is at $y_0 = 0.5$~m in both cases.
    Despite the different physical models, the interface complexity is
    comparable: $D = 1.37$ (MD) versus $D = 1.21$ (continuum).}
\end{figure}

\subsection{Three-dimensional grid convergence}
\label{sec:3d}

\begin{figure}[t]
  \includegraphics[width=\columnwidth]{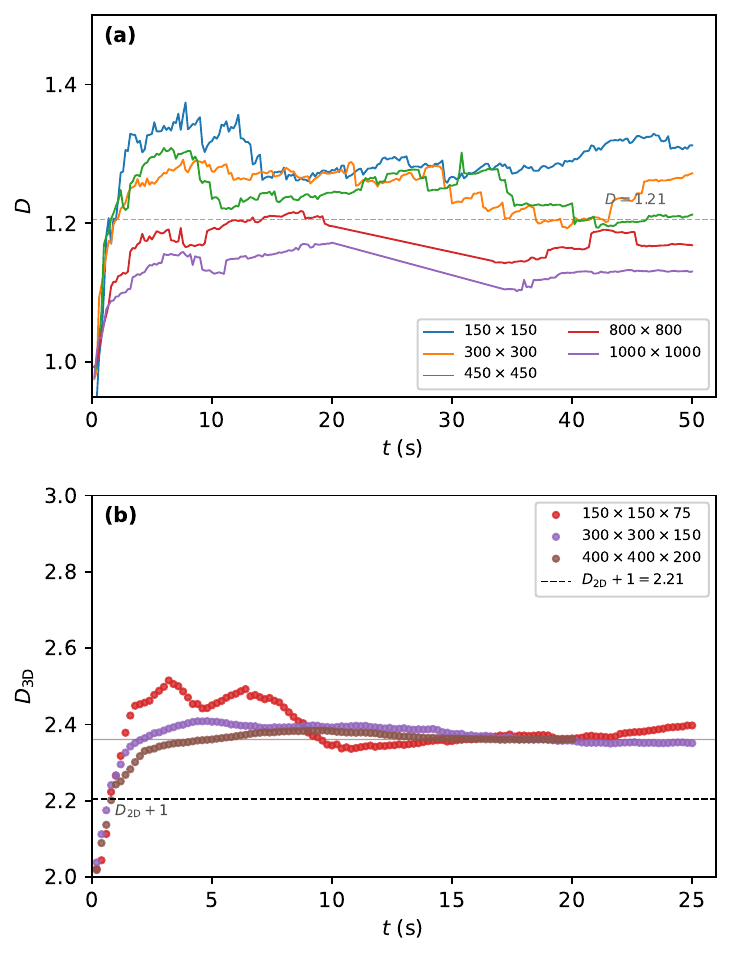}
  \caption{\label{fig:D3d}%
    (a)~Grid convergence of the 2D fractal dimension $D(t)$ for MPLIC at
    three resolutions ($150^2$--$450^2$). The frozen dimension $D = 1.21$ is reached by
    $t \approx 15$~s at $450 \times 450$. (b)~Grid convergence of
    $\Dthree(t)$ for three 3D resolutions. $\Dthree$ freezes near $2.36$
    at all three grids, with the freezing time increasing with resolution.}
\end{figure}

Table~\ref{tab:3d_convergence} presents the 3D grid convergence.
Figure~\ref{fig:D3d} shows the time evolution of $\Dthree$ at all three
resolutions alongside the 2D convergence.

\begin{table}[b]
  \caption{\label{tab:3d_convergence}%
    Grid convergence of the 3D fractal dimension for the velocity-driven
    mixing interface. $\Dthree$ and $\Dproj$ are averaged over the frozen
    portion of the interface evolution at each resolution (see text).
    The scaling range is ${\sim}1$ decade in 3D
    (Sec.~\ref{sec:boxcounting}).
    $\Dtwo + 1 = 2.205$ (using $\Dtwo = 1.205$ from the $450^2$ original IC).}
  \begin{ruledtabular}
  \begin{tabular}{lcccc}
    Resolution & Cells & Window & $\Dthree$ & $\Dproj$ \\
    \hline
    $150 \times 150 \times 75$
      & $1.69 \times 10^6$ & $t = 14$--$25$ & $2.371 \pm 0.011$ & $1.582$ \\
    $300 \times 300 \times 150$
      & $13.5 \times 10^6$ & $t = 20$--$25$ & $2.352 \pm 0.001$ & $1.505$ \\
    $400 \times 400 \times 200$
      & $32 \times 10^6$   & $t = 14$--$20$ & $2.361 \pm 0.002$ & $1.456$ \\
  \end{tabular}
  \end{ruledtabular}
\end{table}

The 3D dimension is nearly resolution-independent, with
$\Dthree \approx 2.36$ at all three grids. At $150 \times 150 \times 75$ ($\Delta x = 6.67$~mm),
$\Dthree = 2.371$; at $300 \times 300 \times 150$, $\Dthree = 2.352$;
at $400 \times 400 \times 200$ ($\Delta x = 2.5$~mm),
$\Dthree = 2.361$. The Marstrand prediction
$\Dtwo + 1 = 2.205$ (using $\Dtwo = 1.205$ from the $450^2$ original IC)
lies $7\%$ below the measured $\Dthree$; as discussed in
Sec.~\ref{sec:ic_sensitivity}, this reflects the IC-realisation
dependence of the 2D dimension at $A = 0$.

The convergence behaviour differs qualitatively from that in 2D. In 2D,
the interface reaches a frozen state by $t \approx 10$~s at all
resolutions, and $D$ is time-independent thereafter. In 3D, different
grids freeze at different times, but the frozen $\Dthree$ shows little
resolution dependence. At $150 \times 150 \times 75$, $\Dthree$ settles to
${\sim}2.36$ by $t \approx 10$~s; at $300 \times 300 \times 150$, the decline
continues to $t \approx 20$~s before stabilising near $2.35$; at
$400 \times 400 \times 200$, $\Dthree$ reaches its plateau of $2.36$ by
$t \approx 14$~s. Averaging over the frozen
portion of each case (Table~\ref{tab:3d_convergence}) therefore requires
resolution-dependent time windows.

The physical origin of the resolution-dependent freezing time is the viscous decay
of the velocity field. The smallest eddies resolved by the grid have
scale $\ell \sim 2\Delta x$ and viscous decay time
$\tau_\nu \sim \ell^2/\nu$. At $150 \times 150 \times 75$, $\tau_\nu \sim (13~\text{mm})^2 /
10^{-5} \approx 17$~s, so the grid-scale eddies have largely decayed by
$t = 10$~s. At $400 \times 400 \times 200$, $\tau_\nu \sim (5~\text{mm})^2 / 10^{-5}
\approx 2.5$~s, and these eddies should have decayed by
$t \sim 5\tau_\nu \approx 12$~s --- consistent with the flattening of
$\Dthree$ at $t \approx 13$--$15$~s.

The projected dimension $\Dproj$ provides a convergence check.
$\Dproj$ converges from above, falling from $1.58$ at $150 \times 150 \times 75$
through $1.51$ at $300 \times 300 \times 150$ to $1.46$ at $400 \times 400 \times 200$.
The grid-converged 3D frozen dimension is
$\Dthree = 2.361 \pm 0.002$ ($400 \times 400 \times 200$, $t = 14$--$20$~s).

\subsection{Marstrand's codimension test}
\label{sec:marstrand}

Table~\ref{tab:marstrand} shows the Marstrand $+1$ test using the
grid-converged values.

\begin{table}[b]
  \caption{\label{tab:marstrand}%
    Marstrand codimension test for the velocity-driven mixing interface,
    using the grid-converged values. The prediction $\Dtwo + 1$ assumes
    isotropic self-similarity between 2D and 3D. The silhouette dimension
    $\Dproj$ is listed for reference only and is not differenced against
    $\Dthree - 1$: projection is not slicing, and it is counted with a
    largest box of $L/2$ rather than the $L/8$ used elsewhere here
    (Sec.~\ref{sec:marstrand}).}
  \begin{ruledtabular}
  \begin{tabular}{lc}
    Quantity & Value \\
    \hline
    $\Dtwo$ (MPLIC $450^2$, $t = 40$--$50$) & 1.205 \\
    $\Dtwo + 1$ (Marstrand prediction) & 2.205 \\
    $\Dthree$ (MPLIC $400 \times 400 \times 200$, $t = 14$--$20$) & $2.361 \pm 0.002$ \\
    Agreement with prediction & $+7\%$ \\
    \hline
    $\Dproj$ (silhouette, $400 \times 400 \times 200$) & $1.456 \pm 0.006$ \\
  \end{tabular}
  \end{ruledtabular}
\end{table}

The $+1$ test shows a $7\%$ discrepancy: $\Dthree = 2.361$ exceeds the
prediction $\Dtwo + 1 = 2.205$ by $0.156$. This discrepancy is
resolution-independent --- $\Dthree$ is flat across grids
(Table~\ref{tab:3d_convergence}) --- and reflects the
IC-realisation dependence of $\Dtwo$ at $A = 0$
(Sec.~\ref{sec:ic_sensitivity}): the 2D and 3D simulations use different
random velocity fields, and the frozen $D$ depends on the realisation.

The projected dimension $\Dproj = 1.46$ at $400 \times 400 \times 200$ converges
from above (from $1.58$ at $150 \times 150 \times 75$). It is reported as a
silhouette measurement and not as a test of the codimension rule, for two
reasons. Projection is not slicing: for a set of dimension above two
Marstrand's projection theorem predicts a shadow of full dimension two, so
$\Dthree - 1 = 1.361$ is not the prediction for $\Dproj$, and the two
should not be differenced~\cite{Douglass2026b}. And $\Dproj$ is counted
with a largest box of $L/2$, where every other dimension reported here uses
$L/8$, so part of any difference between them is a difference of method.
Recounting the silhouette at $L/8$ gives $1.691 \pm 0.012$ over the same
window, which moves it away from $\Dthree - 1$ rather than towards it, as
the projection theorem requires.
The identical analysis pipeline recovers $+1$ to within $1\%$ at matched
mesh spacing for
buoyancy-driven Rayleigh--Taylor
interfaces~\cite{Douglass2026a,Douglass2026b}, where an attractor
eliminates IC dependence.

\begin{figure}[!tb]
  \includegraphics[width=0.85\columnwidth]{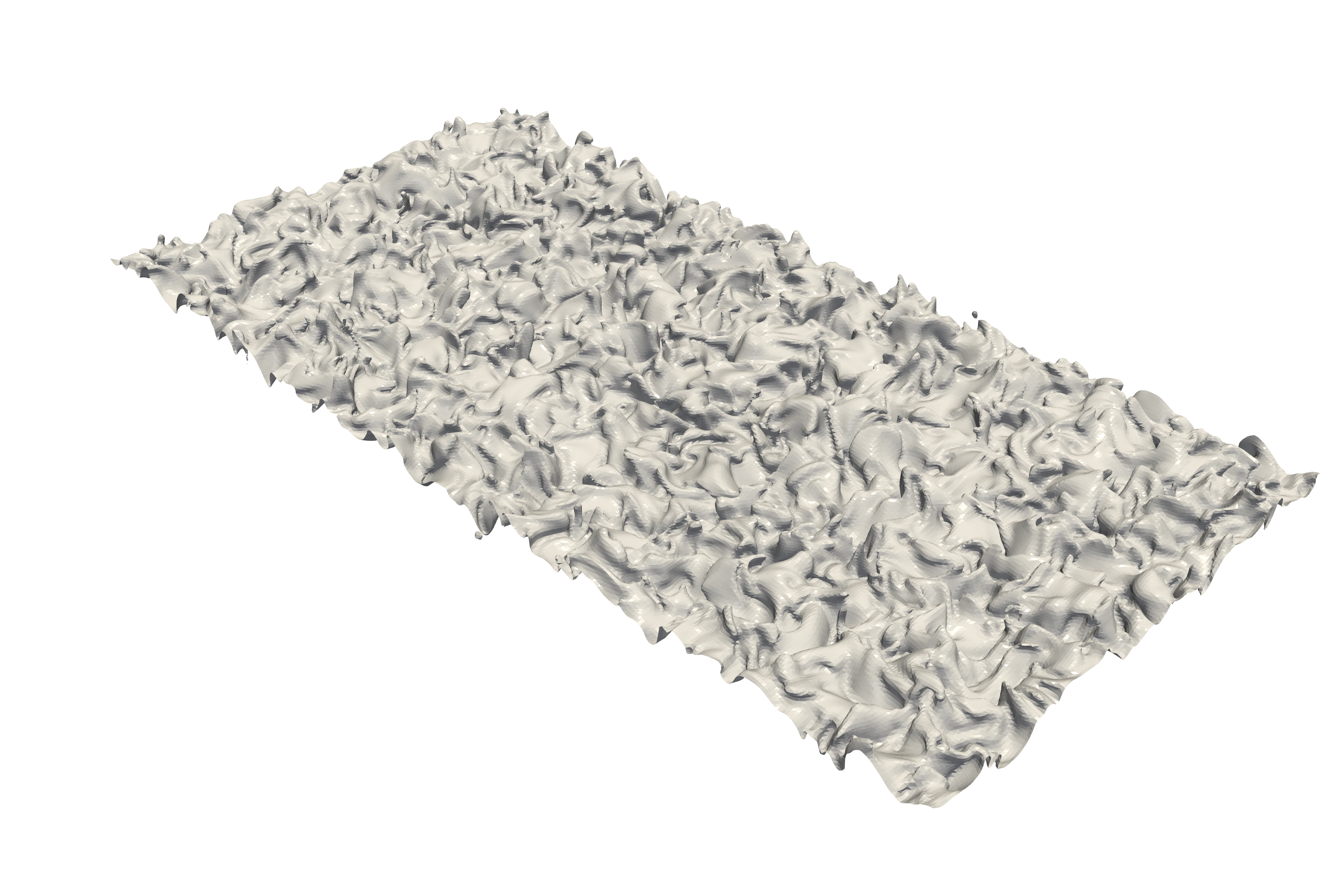}
  \caption{\label{fig:iso}%
    The $F = 0.5$ isosurface at $t = 15$~s in the
    $400 \times 400 \times 200$ simulation ($1.0 \times 1.0 \times 0.5$~m
    domain, $32 \times 10^6$ cells). The interface was initially flat at
    $y_0 = 0.5$~m and has been deformed by the decaying velocity field into
    the frozen configuration shown. The finer grid resolves smaller-scale
    folds and pockets than the coarser grids, yielding
    $\Dthree = 2.361 \pm 0.002$.}
\end{figure}


\subsection{Initial-condition sensitivity and the absence of a fractal attractor}
\label{sec:ic_sensitivity}

The grid convergence in Table~\ref{tab:convergence} conflates two
effects.  The band-limited noise IC (Sec.~\ref{sec:continuum}) is
generated by filtering white noise on an $N_x \times N_y$ array; each
resolution therefore receives a different random velocity field, even
with the same random seed.  The apparent convergence
$D = 1.31 \to 1.24 \to 1.21$ could reflect genuine grid refinement, a
change in the IC realisation, or both.

Extending the series to $600^2$, $800^2$, and $1000^2$
resolves the ambiguity.  Rather than converging further, the frozen
dimension drops to $D = 1.16$ at $600^2$, then to $1.17$ and $1.13$
at $800^2$ and $1000^2$.
The total decline of $|\Delta D| = 0.08$ between $450^2$ and $1000^2$ is
inconsistent with grid refinement; each resolution produces a different
flow and a different frozen interface.  The total geometric mixing width
$h_\mathrm{geo}$ at $t = 50$~s declines nearly fivefold across the
series, from $0.32$~m at $150^2$ to $0.07$~m at $1000^2$
(Table~\ref{tab:ic_sensitivity}), confirming that each grid generates
a qualitatively different velocity pattern.

Two controlled experiments isolate the IC effect from grid refinement.

\textit{Interpolated IC.}  The $450 \times 450$ velocity field was
bilinearly interpolated onto the $600^2$, $800^2$, and $1000^2$ grids, so that
finer grids begin from the same large-scale flow (with some
kinetic-energy loss from smoothing below the source grid scale).  The
frozen dimensions are $D = 1.21$ ($600^2$), $1.21$ ($800^2$), and $1.21$
($1000^2$): the series converges to $D \approx 1.21$.  The mixing width
$h_\mathrm{geo} \approx 0.09$--$0.10$~m matches the source value
($0.09$~m), confirming that the physical flow is preserved.

\textit{Fixed-Fourier IC.}  A single set of $4284$ complex Fourier
coefficients was generated once (seed~42, wavenumber band
$2 \le |\bm{n}| \le 37$, wavelengths $27$--$500$~mm, ensuring
$\ge 4$ cells per wavelength on the coarsest $150^2$ grid) and placed
into each grid's FFT array before inverse-transforming.  Every
resolution sees the identical physical velocity field.  The frozen
dimensions are $D = 1.30$ ($150^2$), $1.32$ ($300^2$), $1.35$ ($450^2$),
$1.37$ ($600^2$), $1.38$ ($800^2$), and $1.38$ ($1000^2$), converging
monotonically toward $D \approx 1.38$ (Table~\ref{tab:ic_sensitivity}).  The mixing
width holds between $0.30$ and $0.33$~m, confirming that
the physics is grid-independent.

\begin{table}[t]
  \caption{\label{tab:ic_sensitivity}%
    Frozen fractal dimension $D$ (averaged $t = 40$--$50$~s) under three
    initial-condition protocols.  Original: independent band-limited
    noise at each grid.  Interpolated: the $450^2$ velocity field
    bilinearly interpolated onto finer grids.  Fixed-Fourier: identical
    Fourier coefficients placed into each grid's FFT array.
    $h \equiv h_\mathrm{geo}$: total geometric mixing width at $t = 50$~s, between the $0.01$
    and $0.99$ levels of the horizontally averaged volume-fraction
    profile $\langle F\rangle(y)$.}
  \begin{ruledtabular}
  \begin{tabular}{lcccccc}
    & \multicolumn{2}{c}{Original}
    & \multicolumn{2}{c}{Interpolated}
    & \multicolumn{2}{c}{Fixed-Fourier} \\
    Grid & $D$ & $h$ (m) & $D$ & $h$ (m) & $D$ & $h$ (m) \\
    \hline
    $150^2$  & 1.31 & 0.32 & ---  & ---  & 1.30 & 0.31 \\
    $300^2$  & 1.24 & 0.15 & ---  & ---  & 1.32 & 0.30 \\
    $450^2$  & 1.21 & 0.09 & ---  & ---  & 1.35 & 0.31 \\
    $600^2$  & 1.16 & 0.09 & 1.21 & 0.09 & 1.37 & 0.31 \\
    $800^2$  & 1.17 & 0.07 & 1.21 & 0.10 & 1.38 & 0.33 \\
    $1000^2$ & 1.13 & 0.07 & 1.21 & 0.10 & 1.38 & 0.33 \\
    \hline
    MD (re-extracted) & \multicolumn{2}{c}{1.37} & & & & \\
  \end{tabular}
  \end{ruledtabular}
\end{table}

At $1000^2$, the three IC protocols produce $D = 1.13$, $1.21$, and
$1.38$ --- a spread of $0.25$, far exceeding the grid-convergence
uncertainty within any single protocol ($\sigma_D \le 0.03$, and
$\le 0.01$ at every grid but one).  The
mixing width tracks $D$: $h_\mathrm{geo} = 0.07$~m (original),
$0.10$~m (interpolated), $0.33$~m (fixed-Fourier).  Broader mixing
zones produce more complex frozen interfaces.

\begin{figure}[b]
  \includegraphics[width=\columnwidth]{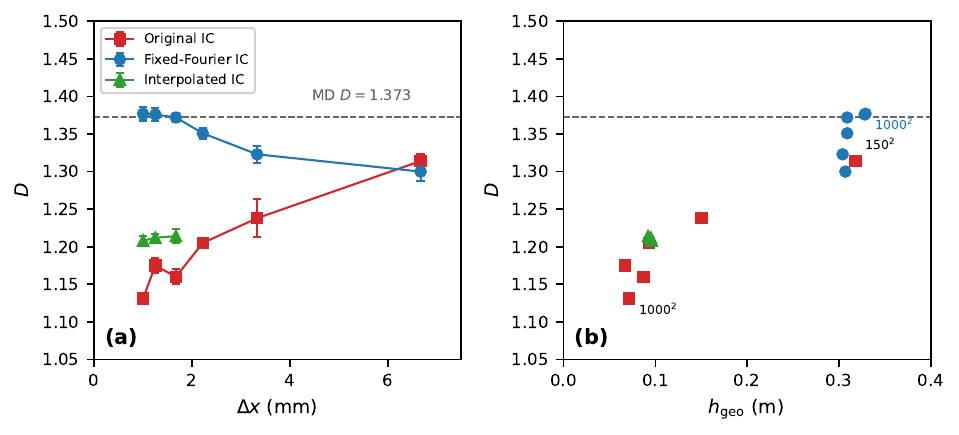}
  \caption{\label{fig:ic_sensitivity}%
    (a)~Frozen fractal dimension $D$ versus grid spacing $\Delta x$ for
    the three IC protocols: original band-limited noise (circles),
    interpolated (squares), and fixed-Fourier (triangles).  The dashed
    line marks the re-extracted MD value $D = 1.37$.  The original-IC
    series decreases because each finer grid
    draws a different random realisation, changing the flow rather
    than refining it.
    (b)~$D$ versus total geometric mixing width $h_\mathrm{geo}$
    at $t = 50$~s.  Across all three protocols, $D$ tracks
    $h_\mathrm{geo}$: broader mixing zones produce more complex
    frozen interfaces.}
\end{figure}

Only the fixed-Fourier experiment constitutes a genuine grid
convergence study; the original IC changes the physics with each
refinement.  Even so, the converged $D \approx 1.38$ depends on which Fourier
coefficients are drawn; a different seed produces a different flow
and a different frozen dimension.

\textit{Absence of an attractor.}  This IC dependence contrasts with
buoyancy-driven Rayleigh--Taylor instability at $A > 0$, where the
fractal dimension converges to $D \approx 1.71$ regardless of the
initial perturbation method, amplitude, or random
seed~\cite{Douglass2026a}.  In the RT case, the instability mechanism
--- buoyancy-driven bubble competition and merger cascade ---
continuously restructures the interface, erasing the memory of the
initial condition.  The fractal dimension acts as an invariant of a
dynamical attractor.

At $A = 0$, no attractor exists.  Without buoyancy, the velocity field
decays monotonically; there is no sustained energy source to
restructure the interface toward a universal geometry.  The frozen
interface is a fossil of the specific velocity field that created it,
and different realisations freeze with different complexity.  The
distinction is fundamental. Any nonzero $A$, however small, activates
the instability mechanism that drives $D$ toward its attractor value,
while $A = 0$ exactly is a qualitatively different physical regime
in which the velocity-driven mixing preserves the IC imprint
permanently.

\textit{Implications for the cross-model comparison.}  The
``grid-converged'' value $D = 1.21$ in Table~\ref{tab:convergence} is
the frozen dimension for one particular IC realisation; it is not a
universal property of $A = 0$ mixing.  Independent LAMMPS simulations show
that the MD dimension ($D \approx 1.37$) measures thermal scatter of atomic positions,
not interface corrugation (Sec.~\ref{sec:universality}).  The
cross-model comparison is therefore categorical --- the two $D$ values
measure different physical quantities --- and the inter-realisation
spread ($D = 1.13$--$1.38$) is a property of the continuum interface
alone.

\textit{Implications for the Marstrand test.}  The codimension test
(Sec.~\ref{sec:marstrand}) uses independent IC realisations in 2D and
3D, and the $7\%$ discrepancy is consistent with the inter-realisation
spread of $D$. The $+1$ rule requires only that the velocity
statistics be isotropic --- a property enforced by the IC construction
--- not that the specific flow pattern be identical. A definitive test
at $A = 0$ would require matching the IC realisation between
2D and 3D simulations.


\section{Discussion}
\label{sec:discussion}

\subsection{Universality of the codimension rule}
\label{sec:universality_codim}

Table~\ref{tab:mechanism} collects the Marstrand $+1$ test across all
interface-generating mechanisms for which both 2D and 3D data are
available.

\begin{table}[b]
  \caption{\label{tab:mechanism}%
    Marstrand $+1$ test across interface-generating mechanisms. Every
    row pairs each dimensionality at its own finest grid, so all three are
    unmatched comparisons and are directly comparable with one another. At
    matched mesh spacing the RT data give $0.6$--$4.5\%$, and $1\%$ or
    less at the four Atwood numbers of Ref.~\cite{Douglass2026b}.
    RT data from Ref.~\cite{Douglass2026a}; passive scalar
    data from Sreenivasan~\cite{Sreenivasan1991}; velocity-stirred data from
    this work. $^*$Sreenivasan's scalar interfaces are outer
    turbulent/non-turbulent boundaries (cross-sectional slices of 3D
    flows, not 2D simulations); Ref.~\cite{Douglass2026a}
    verifies that $D$ is unchanged at $F = 0.1$, $0.5$, and $0.9$,
    closing this threshold caveat. $^\dagger$IC-realisation-dependent
    (Sec.~\ref{sec:ic_sensitivity}); $\Dtwo$ shown is for the
    original $450^2$ IC. $^\ddagger$The upper end of the RT $\Dthree$
    range is from a run that began without its interface perturbation and
    is being recomputed~\cite{Douglass2026a}; that row will move slightly
    when it lands.}
  \begin{ruledtabular}
  \begin{tabular}{lccc}
    Mechanism & $\Dtwo$ & $\Dthree$ & $+1$ deviation \\
    \hline
    RT instability ($A = 0.0021$)$^\ddagger$ & 1.68--1.73 & 2.60--2.62 &
      $\checkmark\;2.8$--$3.2\%$ \\
    Passive scalar (turbulence)$^*$ & 1.36 & 2.35 &
      $\checkmark\;1\%$ \\
    Velocity-stirred ($A = 0$)$^\dagger$ & 1.21 & 2.36 &
      $7\%$ \\
  \end{tabular}
  \end{ruledtabular}
\end{table}

The codimension rule is satisfied by the RT and passive-scalar
mechanisms ($2.8$--$3.2\%$ and $1\%$ on this unmatched pairing, and
$0.6$--$4.5\%$ for RT once the two dimensionalities are compared at
matched mesh spacing), and the velocity-stirred case shows $7\%$. The three
cases span a wide range of fractal dimensions ($\Dtwo = 1.21$ to $1.73$;
the coincidence of Sreenivasan's passive-scalar $\Dtwo = 1.36$ with the MD
thermal-scatter dimension is noted in Sec.~\ref{sec:universality}).
For the velocity-stirred case, the $7\%$ offset reflects the
IC-realisation dependence of $\Dtwo$ at $A = 0$
(Sec.~\ref{sec:ic_sensitivity}): the 2D value $\Dtwo = 1.21$ is the
frozen dimension for one particular random velocity field, while the 3D
value $\Dthree = 2.36$ was generated by a different realisation. Without
an attractor to drive both toward a common dimension, the $+1$ test
conflates the codimension relationship with inter-realisation scatter.

\begin{figure}[!tb]
  \includegraphics[width=0.9\columnwidth]{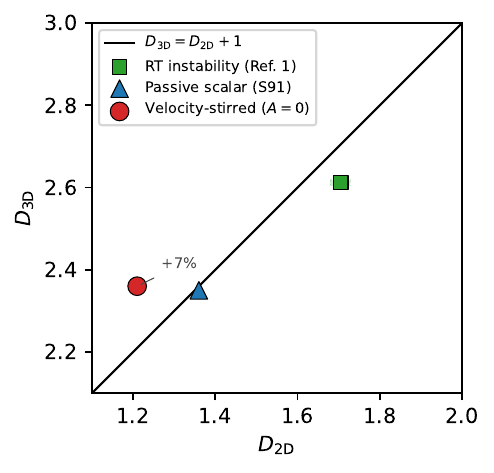}
  \caption{\label{fig:mechanism}%
    Marstrand $+1$ test across interface-generating mechanisms. The
    diagonal line is $\Dthree = \Dtwo + 1$. RT instability (square with
    shaded range, $A = 0.0021$; data from
    Ref.~\cite{Douglass2026a}), Sreenivasan's passive
    scalar (triangle), and the velocity-stirred interface (circle, this
    work). RT and passive-scalar points lie on the $+1$ line;
    the velocity-stirred point sits $7\%$ above it, reflecting
    IC-realisation dependence at $A = 0$.}
\end{figure}

The velocity-stirred case provides a stringent test. In three
dimensions, vortex stretching transfers kinetic energy from large scales to
small scales in a forward (Kolmogorov) cascade, generating fine-scale
interface wrinkles across the full inertial range. In two dimensions, the
absence of vortex stretching reverses the cascade
direction~\cite{Kraichnan1967}: energy flows to large scales (inverse
cascade), while enstrophy cascades forward with a steeper $k^{-3}$
spectrum. A 2D simulation of velocity-driven mixing is therefore a
different dynamical system from its 3D counterpart, and
the cascade asymmetry could, in principle, produce different interface
geometries in 2D and 3D.

The $7\%$ discrepancy in the velocity-stirred case does not invalidate the
codimension relationship but reveals a confounding factor: at $A = 0$, the
frozen $D$ depends on the specific IC realisation
(Sec.~\ref{sec:ic_sensitivity}), and the 2D and 3D simulations used
different random velocity fields. The frozen interface inherits the
isotropy of the velocity field that created it --- and since Marstrand's
theorem requires only isotropic self-similarity, the $+1$ relationship
should hold within a single realisation. The measured discrepancy is of
the same order as the inter-realisation spread ($\Delta D \approx 0.32$
across IC protocols), suggesting that it is an IC effect rather than a
failure of isotropy.

This argument explains the $+1$ results for the three mechanisms
in Table~\ref{tab:mechanism}:
\begin{itemize}
  \item \textit{Buoyancy-driven instability} (RT): the bubble--merger
    mechanism is dimension-invariant and generates isotropic interface
    geometry at all scales. The $+1$ deviation is $2.8$--$6.5\%$ on an
    unmatched comparison and $0.6$--$4.5\%$ at matched mesh spacing,
    reaching $1\%$ or less there at the four higher Atwood numbers, and it
    falls as the three-dimensional grid is refined.
  \item \textit{Passive scalar} (Sreenivasan): the 2D measurement is
    a literal Marstrand slice of the 3D scalar field; $+1$ holds by the
    theorem's premise.
  \item \textit{Velocity-stirred} (this work): the cascade direction
    differs between 2D and 3D, and the statistical isotropy of the velocity
    field is preserved, but the absence of an attractor means the $+1$
    test is confounded by IC-realisation scatter ($7\%$).
\end{itemize}

\subsection{The coarse-grid artefact}
\label{sec:artefact}

The $+1$ discrepancy ($7\%$ at all grids) differs from the buoyancy-driven
RT case, but not in the way a comparison of magnitudes suggests. There the
unmatched deviation is $2.8$--$6.5\%$, which overlaps the value measured
here, and it \emph{falls} as the three-dimensional grid is refined, from
$5.4$--$6.5\%$ at the coarsest grid to $3.3$--$3.7\%$ at the finest, reaching
$0.6$--$4.5\%$ once the test is made at matched mesh spacing and $1\%$ or
less there at the four higher Atwood
numbers~\cite{Douglass2026a,Douglass2026b}. The velocity-stirred discrepancy
does none of that. It is $7\%$ at every grid, with $\Dthree$ flat at
$2.371$, $2.352$ and $2.361$ across a factor of nineteen in cell count. That
is the distinction, and it is a behavioural one rather than a difference of
size. Two effects contribute to the velocity-stirred case's larger error:

First, the 3D scaling range is only ${\sim}1$ decade
(Sec.~\ref{sec:boxcounting}), which produces substantial per-snapshot scatter
in $\Dthree$ [Fig.~\ref{fig:D3d}(b)]. This scatter is symmetric
and does not bias the frozen-window averages, but it inflates the
measurement uncertainty relative to the 2D case.

Second, the IC-realisation dependence (Sec.~\ref{sec:ic_sensitivity}):
the 2D and 3D simulations use different random velocity fields. At
$A > 0$, the attractor drives both toward a common $D$, making the $+1$
test insensitive to IC differences. At $A = 0$, the frozen $D$ retains
the IC imprint, and the $+1$ test measures the distance between two
independent realisations in addition to any true geometric relationship.

The resolution-dependent freezing time is specific to decaying
velocity-driven mixing. In buoyancy-driven RT,
the instability mechanism is continuously active and the interface reaches a
statistical steady state that is resolution-independent at converged grids.
There the $+1$ error shrinks with refinement and reaches $1\%$ or less at
matched mesh spacing~\cite{Douglass2026a,Douglass2026b}, whereas here it is
unchanged from the coarsest grid to the finest.

\subsection{What does the MD dimension measure?}
\label{sec:universality}

While the $+1$ test probes geometric universality \textit{across}
dimensions, the original question of Cao \textit{et al.}\ --- whether the
fractal dimension depends on the physical model --- concerns universality
\textit{within} a fixed dimension.  To test this, we reproduced the MD
mixing in LAMMPS at system sizes from $12\,000$ to $500\,000$ atoms,
using the same $6$--$12$ Lennard-Jones interaction as Cao
\textit{et al.}~\cite{Cao1992}, who state an atomic mass of $16$~a.u.\
chosen to accelerate their simulation~\cite[p.~3841]{Cao1992}.

Box counting the midpoint cloud at the atomic scale ($\ell =
0.8$--$2.5\sigma$) gives $D \approx 1.36$ for every system size,
every temperature, and every timestep --- including $t = 0$, when the
interface is flat by construction ($D = 1.35$).  Late-time averages
are $D = 1.35$--$1.37$ across all five system sizes (12k--500k atoms),
with standard deviations of $0.003$--$0.009$; the dimension is
converged from the smallest system.  Cao \textit{et al.}'s temperature
insensitivity --- their Fig.~2(c) shows 130, 300 and 900~K converging to
the same late-time $D$ --- is consistent with ours over the narrower
$T^* = 1.0$--$5.0$ we ran, since the ratio of thermal displacement to
lattice spacing changes only modestly across either range.

\begin{figure*}
  \includegraphics[width=\textwidth]{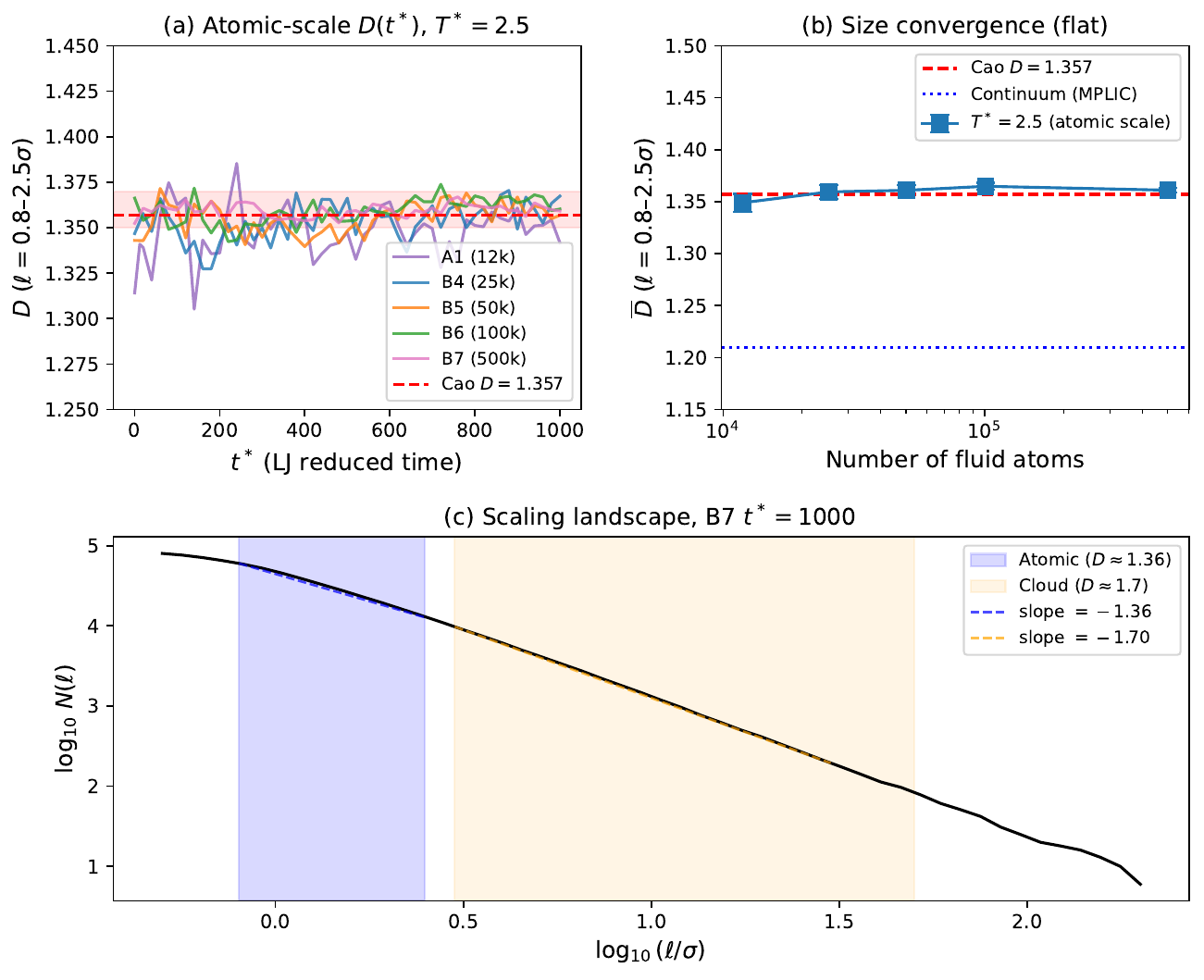}
  \caption{\label{fig:md_scaling}%
    LAMMPS molecular dynamics results (500k-atom case, $T^* = 2.5$).
    (a)~Atomic-scale box-counting dimension $D(\tau)$ for five system sizes
    (12k--500k atoms).  $D \approx 1.36$ at all times,
    including $\tau = 0$ when the interface is flat; the red dashed line
    marks Cao's published value $D = 1.357$.
    (b)~Late-time $\bar{D}$ versus number of fluid atoms, confirming
    convergence from the smallest system.  The dotted blue line marks
    the continuum MPLIC value $D = 1.21$.
    (c)~Box-counting log-log curve at $\tau = 1000$.
    Two regimes are visible: at $\ell =
    0.8$--$2.5\sigma$ (blue shading), the slope gives $D \approx 1.36$,
    measuring thermal scatter of atomic positions; at $\ell =
    3$--$50\sigma$ (orange shading), the slope steepens to $D \approx
    1.7$, reflecting the filling of the midpoint cloud.
    (d)~Scale dependence of $D(\tau)$.  The atomic-scale dimension
    (blue) is flat at $1.36$ from $\tau = 0$ onward; the mesoscale
    dimension (orange) rises from $1.59$ to $1.71$, capturing the
    increasing interface corrugation but starting far above $D = 1.0$
    (grey dotted line) due to thermal contamination of the midpoint
    cloud even when the interface is flat.}
\end{figure*}

In the continuum simulations, a flat interface at $t = 0$ gives
$D = 1.0$; that the MD value differs ($D = 1.35$) establishes that
atomic-scale box counting does not measure interface geometry.  The
dimension is instead set by the thermal scatter of discrete atomic
positions about their lattice sites.  Cao \textit{et al.}\
recognised the difficulty, noting that at $t \approx 0$ the midpoint
convention ``did not produce an exactly straight line'' and that $D$ was
``essentially arbitrary'' at early times~\cite{Cao1992}.  They argued
that at late times, when the interface structure spans 5--10
interatomic spacings, the ${\sim}1$-spacing thermal uncertainty becomes
negligible and $D$ is insensitive to the surface definition.  Our data
show that this argument is incomplete.  Box counting at the atomic scale
($\ell = 0.8$--$2.5\sigma$) gives $D \approx 1.36$ regardless
of the mixing state, so the measurement remains dominated by thermal
scatter even when the physical interface structure is much larger.  The
box-counting curve for the $500\,000$-atom case makes the scale
separation explicit [Fig.~\ref{fig:md_scaling}(c)].  At $\ell =
0.8$--$2.5\sigma$, the slope gives $D \approx 1.36$ (measuring the
atomic point spread), while at $\ell = 3$--$50\sigma$, the slope
steepens to $D \approx 1.7$, reflecting the filling of the midpoint
cloud as its $y$-extent grows with the mixing width.

The two regimes respond differently to the evolving interface.
Figure~\ref{fig:md_scaling}(d) shows $D(\tau)$ measured at both scales
for the $500\,000$-atom case.  At the atomic scale, $D$ is flat at
${\approx}1.36$ from $\tau = 0$ to $\tau = 1000$, confirming that it
records thermal scatter rather than interface geometry.  At the
mesoscale, $D$ rises from $1.59$ at $\tau = 0$ to $1.71$ at late
times, capturing the increasing corrugation that is visually evident
in Cao \textit{et al.}'s Figs.~1(a) and 3(b).  But even the
mesoscale value at $\tau = 0$ --- when the interface is flat by
construction --- is $1.59$, far above the $D = 1.0$ of a smooth line.
The midpoint cloud is a two-dimensional point set whose thickness
(${\sim}60\sigma$ of thermal scatter even at $\tau = 0$;
Fig.~\ref{fig:md_scaling}d) contaminates the box-counting slope at
all accessible scales.  The mesoscale $D$
thus conflates two contributions: a thermal pedestal ($D = 1.59$ at
$\tau = 0$, when the interface is geometrically flat) and an
advective increment ($\Delta D \approx 0.12$, from the velocity
field deforming the interface).  By contrast, the grid-converged
continuum dimension ($D = 1.21$) measures only the advective
contribution, starting from $D = 1.0$.  It is tempting to subtract
the thermal baseline and compare $1.71 - 0.59 = 1.12$ with $1.21$,
but the two contributions are entangled in box counting --- the
thermal scatter changes the \emph{effective topology} of the point
set, not merely its amplitude --- so the subtraction is not
well-defined.  Neither regime measures the fractal dimension of the
mixing interface in the continuum sense.

\begin{figure}
  \includegraphics[width=\columnwidth]{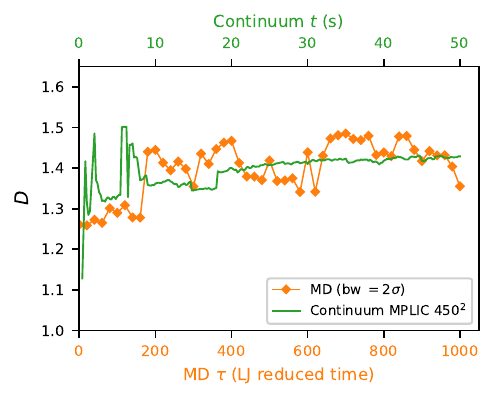}
  \caption{\label{fig:md_interface}%
    Fractal dimension trajectories from the MD simulation (orange) and
    the continuum MPLIC $450^2$ simulation (green), plotted on
    independent time axes (bottom: LJ reduced time; top: seconds).
    The MD curve is obtained by reducing the midpoint cloud to a
    one-dimensional curve (median $y$ per $x$-bin, bin width $=
    2\sigma$, $\approx 55$ atoms per bin) and box-counting the
    resulting segments.  Both trajectories start near $D \approx 1.2$
    and converge to $D \approx 1.43$.  The MD curve's $\tau = 0$
    baseline of $1.26$ (vs.\ the ideal $1.0$) reflects residual thermal
    scatter surviving the median binning; late-time MD values oscillate
    within $\pm 0.05$ of the continuum.}
\end{figure}

A more direct comparison is possible by reducing the midpoint cloud
to a one-dimensional curve: binning in $x$ and taking the median $y$
per bin (bin width $= 2\sigma$, ${\approx}\,55$ atoms per bin)
suppresses most thermal scatter while preserving the large-scale
corrugation.  Box counting the resulting segments yields the
trajectory shown in Fig.~\ref{fig:md_interface}: the MD and
continuum curves both start near $D \approx 1.2$ and converge to
$D \approx 1.43$.  The residual $\tau = 0$ baseline of $1.26$
(vs.\ the ideal $1.0$) reflects thermal scatter surviving the median
filter; late-time MD values oscillate within $\pm 0.05$ of the
continuum, consistent with the binning noise floor.

At the atomic scale, the cross-model comparison is categorical, not
quantitative: the continuum value $D = 1.21$ records interface
corrugation, while the MD value $D = 1.36$ records the spatial
distribution of atoms vibrating about lattice positions.  These are
different physical quantities.  However, when the midpoint cloud is
reduced to a one-dimensional curve by median binning
(Fig.~\ref{fig:md_interface}), the MD trajectory converges to
$D \approx 1.43$, the same late-time value as the continuum.  The
remaining discrepancy --- a $\tau = 0$ baseline of $1.26$ rather than
$1.0$ --- is the binning noise floor.  The reduce-to-curve analysis
thus recovers the universality that raw atomic-scale box counting
obscures: the two physical models produce the same interface geometry,
but measuring it from MD requires filtering the thermal scatter first.

The Marstrand $+1$ test
(Sec.~\ref{sec:marstrand}) compares the \textit{same} physical
quantity --- continuum box-counting $D$ --- in two and three
dimensions.  The monofractal character of the velocity-driven
interface ($D_0 \approx D_1 \approx D_2$) is a continuum property
confirmed across all three IC protocols
(Sec.~\ref{sec:ic_sensitivity}).  What the atomic-scale MD analysis
shows is that raw box counting at $\ell < 3\sigma$ measures thermal
noise, not geometry; the interface dimension can only be compared
across physical models after the thermal contribution is removed.

\section{Conclusions}
\label{sec:conclusions}

We have revisited the fractal analysis of Cao \textit{et al.}~\cite{Cao1992}
with three decades of methodological improvement.

\begin{enumerate}
  \item \textit{Reanalysis and IC sensitivity.} Modern geometric VOF
    reconstruction on grids up to $1000 \times 1000$ reveals that the
    frozen continuum dimension at $A = 0$ depends on the velocity IC
    realisation, with converged values of $D = 1.13$, $1.21$, and $1.38$
    under three controlled IC protocols. The IC dependence reflects the
    absence of a fractal attractor at $A = 0$ --- without buoyancy, the
    frozen interface retains the imprint of the initial velocity field
    permanently, in contrast with $A > 0$ RT, where $D$ is near $1.71$ at the
    finest grid whatever the initial-condition family~\cite{Douglass2026a}.  Independent LAMMPS MD simulations (12k--500k
    atoms) show that the MD dimension $D \approx 1.36$ measures
    thermal scatter of atomic positions about lattice sites, not
    interface geometry --- a flat interface at $t = 0$ gives the same $D$
    as a fully mixed domain.  The result is scale-dependent: at the
    mesoscale ($\ell = 3$--$50\sigma$), $D$ does increase with time
    (1.59 to 1.71), capturing the growing interface corrugation, but
    even then a flat interface gives $D = 1.59$, not~$1.0$, because
    the midpoint cloud's thermal thickness contaminates box counting at
    all accessible scales.  At the atomic scale, the cross-model
    ``gap'' is categorical --- the two quantities being compared are
    physically different.  However, reducing the midpoint cloud to a
    one-dimensional curve (median $y$ per $x$-bin at $2\sigma$) and
    box-counting the resulting segments recovers $D \approx 1.43$,
    matching the continuum: the interface geometry is universal across
    physical models once thermal scatter is filtered.

  \item \textit{Theoretical bounds.} The fractal dimension of the
    linear-regime interface is exactly $D = 2 - \beta$ at $A > 0$ and
    $D = \frac{5}{2} - \beta$ at $A = 0$, where $\beta$ is the spectral
    decay exponent of the initial velocity perturbation; the half-unit
    shift is the Lipschitz exponent contributed by the RT growth-rate
    factor $\sqrt{n}$.  For the white-noise IC used by Cao ($\beta = 0$),
    the series does not converge and the theorem makes no prediction.
    The early-time $D \approx 2$ reported by Cao \textit{et al.}\ is a
    box-counting boundary artefact; removing it eliminates the apparent
    continuum/atomistic distinction.

  \item \textit{3D grid convergence.} The three-dimensional
    velocity-driven mixing interface has $\Dthree = 2.361 \pm 0.002$ at
    the finest grid ($400 \times 400 \times 200$), nearly
    resolution-independent across all three grids ($\Dthree \approx 2.36$).
    The Marstrand $+1$ test shows a $7\%$ discrepancy, reflecting
    the IC-realisation dependence of $\Dtwo$ at $A = 0$ rather than
    a failure of isotropy.

  \item \textit{Codimension rule.} Combined with the RT data of
    Refs.~\cite{Douglass2026a,Douglass2026b}, where the unmatched $+1$
    deviation is $2.8$--$6.5\%$ and falls under refinement, and with passive
    scalar mixing ($1\%$), the velocity-stirred case shows a $7\%$
    discrepancy that does not move with resolution and is
    consistent with the inter-realisation spread of $D$ at $A = 0$.
    A definitive test of $+1$ in this regime requires matching the IC
    realisation between 2D and 3D simulations.
\end{enumerate}

\begin{acknowledgments}
  This research received no specific grant from any funding agency,
  commercial or not-for-profit sectors.
  Claude Opus 4.6 (Anthropic, 2026) was used for literature verification,
  drafting assistance, and prose critique; all simulations, analysis code,
  numerical results, figures, and scientific conclusions are the authors'
  own work.
  The authors declare no competing interests.

  \textit{Author contributions.}
  R.W.D.\ conceived and designed the study, performed all continuum and
  molecular-dynamics simulations, developed the analysis code, produced
  all figures, and wrote the manuscript.
  S.R.D.\ derived the theoretical bounds on the interface dimension
  (Sec.~III) and reviewed the manuscript.

  \textit{Data availability.}
  Simulation data and analysis scripts are available from the
  corresponding author upon reasonable request.
\end{acknowledgments}

\bibliography{cao-reanalysis}

\end{document}